\documentclass[reprint,pra,aps,showpacs,letterpaper,superscriptaddress]{revtex4-1}
\usepackage{times}
\usepackage[utf8]{inputenc}
\usepackage[english]{babel}
\usepackage{bm}
\usepackage{graphicx}
\usepackage{amsmath}
\usepackage{amssymb}
\usepackage{amsfonts}
\usepackage{braket}
\usepackage{xcolor}
\usepackage{hyperref}
\hypersetup{
    pdfstartview={FitH},    
    colorlinks=true,       
    linkcolor=blue,          
    citecolor=blue,        
    filecolor=blue,      
    urlcolor=blue           
}

\usepackage{subfigure}

\begin{document}

\title{Relativistic mask method for electron momentum distributions after 
ionization of hydrogen-like ions in strong laser fields}

\author{D.~A.~Tumakov}
\email{dm.tumakov@gmail.com}
\affiliation{Department of Physics, St.~Petersburg State University,
Universitetskaya Naberezhnaya 7/9, St.~Petersburg 199034, Russia}
\author{Dmitry~A.~Telnov}
\email{d.telnov@spbu.ru}
\affiliation{Department of Physics, St.~Petersburg State University,
Universitetskaya Naberezhnaya 7/9, St.~Petersburg 199034, Russia}
\author{G.~Plunien}
\affiliation{Institut f\"ur Theoretische Physik, Technische Universit\"at
Dresden, Mommsenstrasse 13, Dresden D-01062, Germany}
\author{V.~A.~Zaytsev}
\affiliation{Department of Physics, St.~Petersburg State University,
Universitetskaya Naberezhnaya 7/9, St.~Petersburg 199034, Russia}
\author{V.~M.~Shabaev}
\affiliation{Department of Physics, St.~Petersburg State University,
Universitetskaya Naberezhnaya 7/9, St.~Petersburg 199034, Russia}

\begin{abstract}
Wavefunction-splitting or mask method, widely used in the non-relativistic calculations of 
the photoelectron angular distributions, is extended to 
the relativistic domain within the dipole approximation. Since the closed-form 
expressions for the relativistic Volkov states are not available within the 
dipole approximation, we build such states numerically solving a single 
second-order differential equation. We calculate the photoelectron energy 
spectra and angular distributions for highly charged ions under 
different ionization regimes with both the direct and the relativistic mask 
methods. We show that the relativistic mask method works very well and 
reproduces the electron energy and angular distributions calculated by the 
direct method in the energy range where both methods can be used. On the other 
hand, the relativistic mask method can be applied for longer laser pulses 
and/or higher photoelectron energies where the direct method may have 
difficulties.
\end{abstract}
\maketitle
\section{Introduction}\label{sec:intro}
With recent advancement of the laser technologies making 
it possible  to generate extremely intense short pulses, the light-matter 
interaction phenomena draw much attention both in the theory and 
experiment~\cite{brabec2000,agostini2004,krausz2009,dipiazza2012}. The most 
powerful free-electron laser facilities, such as XFEL \cite{tschen2017} at 
Hamburg and LCLS \cite{dunne2017} at Stanford, are expected to produce 
electromagnetic fields with the peak brilliance of up to 5$\times$10$^{33}$ 
photons/s/mm$^{2}$/mrad$^{2}$/0.1\% bandwidth and wavelengths down to 0.05~nm. 
Such extremely strong and high-frequency fields take the electronic dynamics of 
the target to the relativistic domain. Interaction of these fields with 
highly-charged ions is of particular interest, given the fact that such ions 
have their own strong Coulomb fields where the electronic motion is essentially 
relativistic. In this respect, we should mention the shortly upcoming 
High-Intensity Laser Ion-Trap Experiment (HILITE) 
experiment~\cite{vogel2012,ringleb2015,stallkamp2020}, which is intended to 
study the light-matter interaction 
using the Penning trap.

For the correct theoretical description of such processes, a relativistic treatment should be invoked 
to capture the electron dynamics: not only the bound electron is relativistic for the ions with high nucleus charge $Z$, 
but also the ionized electrons can reach the relativistic velocities. Various approaches exist today, including relativistic 
extensions of the strong field approximation (SFA)~\cite{klaiber2013}, numerical solution of the 3D 
Dirac equation~\cite{bauke2011,fg2012,vanne2012,telnov2013,rozenbaum2014,ivanov2015,kjellsson2017,ivanova2018}, 
and the modification of the Schr{\"o}dinger equation 
allowing to account for relativistic effects to a large extent~\cite{lindblom2018}.

Photoelectron angular and energy distributions contain various information 
about both the ionization process and internal structure of 
the target. Besides the familiar above-threshold ionization 
(ATI) peaks~\citep{agos1979} and rescattering plateau~\cite{corkum1993} they 
can feature many subtle effects: ``channel-closing''~\cite{muller1983, 
kruit1983}, Stark-induced Rydberg states resonances (Freeman 
resonances)~\cite{freeman1987,potvliege2009}, low energy structure (LES) 
attributed to the Coulomb focusing effect~\cite{blaga2009,liu2010}, or 
interference structure originating from interfering electrons, emitted at 
different 
times~\cite{telnov1995,lindner2005,wickenhauser2006,arbo2006,arbo2010,arbo20102,
tumakov2019}. Advances in the experimental setups draw a lot of attention to 
these fine structures and effects present in the electron spectra.

For the direct calculation of the photoelectron angular distributions (PAD) in the relativistic regime 
one has to solve the full-dimensional time-dependent Dirac equation numerically. However, this solution is 
computationally expensive, especially if the ionized wavepacket travels a long distance from the ionic core. 
There are a number of non-relativistic well-established methods to extract the PAD without the full 
information about the wave packet in remote regions of the space: the window operator technique~\cite{schafer1990}; 
geometrical splitting of the coordinate space and 
the wavefunction (mask method)~\cite{chelkowski1998,tong2006,giovannini2012}, which can be directly connected to the usage 
of complex absorbing potentials (CAPs) or smooth exterior complex scaling~\cite{giovannini2015}; 
description of the electron dynamics in the Kramers-Henneberger reference frame~\cite{telnov2009,telnov2011}; 
calculation of the flux through a spherical surface placed far enough from the core~\cite{tao2012}, and the 
solution of the dynamical equations in the momentum representation~\cite{zhou2011,zhou2013}.

In this contribution, we introduce the relativistic 
generalization of the mask method for the hydrogen-like ions and provide its 
numerical illustrations studying the  
ionization of such ions by strong linearly-polarized laser pulses. 
Since the method
is based on the Dirac equation, it enables a correct and natural treatment 
of the relativistic electronic structure of highly-charged ions.
The method  
can be used to obtain the PAD without propagation of the wave function to long 
times and large distances for the hydrogen-like system, and can also be directly 
generalized to the many-electron problems within the density functional theory 
framework. The present formulation of the method is restricted to the dipole 
approximation for the interaction of the electron with the electromagnetic 
field. With increasing the field intensity the dipole 
approximation will eventually break down. Depending on the wavelength, the 
applicability of this approximation may become questionable even for the fields 
which are not extremely strong. This issue has been widely discussed in the 
literature (see, for example, 
Refs.~\cite{reiss2001,palaniyappan2006,reiss2008,ludwig2014,klaiber2017}). 
We should mention two different aspects here. First, applicability of the dipole 
approximation is limited to the radiation wavelengths which are much larger than 
the size of the target, so the spatial dependence of the field can be safely 
ignored. The second limitation is related to the influence of the magnetic 
field component on the electronic motion, which is more significant for 
stronger fields and faster electrons. When the nondipole effects become 
important, the method presented in this paper cannot be used for accurate 
evaluation of the photoelectron distributions. However, it still can be used to 
estimate the magnitude of the nondipole corrections. To do so one has to 
compare the dipole and nondipole results, thus the calculations within the 
dipole approximation make sense even in the case when this approximation is not 
very accurate. Still, a wide range of photon energies and field intensities 
exists where the dipole approximation is expected to work reasonably well. In 
our case studies, we use the field parameters where the dipole approximation is 
well justified.

The paper is organized as follows. In Sec.~\ref{sec:theory} we describe in
detail the theoretical and computational methods applied to the present problem. 
The examples of the method implementation, the results of our calculations, and 
all necessary theoretical analyses are presented in Sec.~\ref{sec:results}.
Sec.~\ref{sec:summary} contains the concluding remarks. Atomic units are used 
throughout the paper ($\hbar = m_e = |e| = 1$), unless specified otherwise.

\section{Theoretical and computational methods}\label{sec:theory}
\subsection{Hydrogen-like ion exposed to a strong laser field}\label{subsec:tdde}

Relativistic dynamics of the electron in a hydrogen-like ion is governed by the time-dependent 
Dirac equation (TDDE):
\begin{equation}\label{eq:tdde}
i \frac{\partial}{\partial t} \Psi (\bm{r}, t) = \left[ H_0 + V \left(\bm{r}, t \right) \right] \Psi 
(\bm{r}, t),
\end{equation}
where the stationary part of the Hamiltonian reads
\begin{equation}\label{eq:h0}
H_0 =  c \boldsymbol{\alpha} \cdot \bm{p} + c^2 \beta + V_{\rm nucl},
\end{equation}
with $c \approx$ 137 being the speed of light; $\boldsymbol{\alpha}$ and $\beta$ are the 
Dirac matrices. Spherically-symmetric potential 
$V_{\rm nucl}$ describes the interaction with the nucleus: 
\begin{equation}\label{eq:vnucl}
V_{\rm nucl} = - \frac{Z(r)}{r},
\end{equation}
where $Z(r)$ is the effective nuclear charge. In all our calculations we use 
the model of a charged sphere for the nucleus with 
nuclear radii taken from Ref.~\cite{angeli2013}. We consider the interaction with the external 
electromagnetic field of the laser pulse within the dipole approximation in the 
velocity gauge, so the interaction term reads
\begin{equation}\label{eq:vint}
V(\bm{r}, t) = c \boldsymbol{\alpha} \cdot \bm{A}(t)
\end{equation}
with the ``vector potential'' defined as $\bm{A}(t) = - \int_{-\infty}^t \bm{F}(\tau) d \tau$,
where $\bm{F}(\tau)$ is the electric field strength. We assume the electric field 
to be a linearly-polarized laser pulse along the $z$ axis with sine-squared 
envelope for the vector potential:
\begin{equation}\label{eq:efield}
\bm{A}(t) = 
\begin{cases}
- \dfrac{F}{\omega} \bm{e_z} \sin^2{\dfrac{\omega t}{2 N}} \sin{\omega t} 
 & ( 0
\leqslant t \leqslant T), \\
0 &  (t<0,\ t>T),
\end{cases}
\end{equation}
where $\omega$ and $F$ are the carrier frequency and the peak field 
strength, respectively, $N$ denotes the number of 
optical cycles, and $T = \frac{2 \pi N}{\omega}$ is the pulse duration.

Since the full Hamiltonian is axially-symmetric, the initial projection $m = 1/2$ of the 
electron total angular momentum $j$ on $z$ axis is conserved during the interaction. Utilizing this 
fact we can factor out the dependence on the angle $\varphi$ (rotation angle about the $z$-axis) in 
the wavefunction explicitly, reducing the problem's dimension: 
\begin{equation}\label{eq:wavef2d}
\Psi(\bm{r}, t) = \frac{1}{\sqrt{2 \pi} r} \begin{pmatrix} \psi_1(r, \theta, t) \\ {\rm e}^{i \varphi} \psi_2(r, \theta, t) \\ i \psi_3(r, \theta, t) \\ i {\rm e}^{i \varphi} \psi_4(r, \theta, t) \end{pmatrix}.
\end{equation}
The four-component function $\psi(r, \theta, t)$ formed by the scalar functions 
$\psi_i(r, \theta, t)$ satisfies the following equation:
\begin{equation}\label{eq:tdde2d}
i \frac{\partial}{\partial t} \psi(r, \theta, t) = \widetilde{H}(t) \psi(r, \theta, t),
\end{equation}
where
\begin{equation}
\widetilde{H}(t) = \begin{pmatrix}
\left(V_{\rm nucl} + c^2 \right) \cdot \bm{1}_2  & c \left(D + i A_z(t) \sigma_z\right) \\
-c \left(D + i A_z(t) \sigma_z\right) & \left(V_{\rm nucl} - c^2\right) \cdot \bm{1}_2
\end{pmatrix}
\end{equation}
and
\begin{equation}
  \begin{split}
    D = \left(\sigma_x \sin \theta + \sigma_z \cos \theta \right) 
\left(\frac{\partial}{\partial r} - \frac{1}{r} \right) &       \\
    + \frac{1}{r} \left( \sigma_x \cos \theta - \sigma_z \sin \theta \right) \frac{\partial}{\partial \theta} + &  
\frac{1}{2 r \sin \theta} \left(\sigma_x + i \sigma_y \right)
  \end{split}
\end{equation}
with $\sigma_x$, $\sigma_y$, and $\sigma_z$ being the Pauli matrices, and 
$\bm{1}_2$ being the 2 $\times$ 2 unity matrix.
 
To obtain the initial state for the time-dependent problem~\eqref{eq:tdde2d} we consider first the 
time-independent Dirac equation for the electron in a hydrogen-like ion:
\begin{equation}\label{eq:de2d}
\widetilde{H}(t = 0) \phi_n = \varepsilon_n \phi_n.
\end{equation}
The problem~\eqref{eq:de2d} can be solved numerically with the 
straightforward implementation of the generalized pseudospectral (GPS) method 
(for the details, see, for example, \cite{yao1993,telnov1999,telnov2013}). 
Discretization of the Eq.~\eqref{eq:de2d} leads to the symmetric matrix eigenvalue problem, which can 
be solved efficiently with the linear algebra routines. 

It is well-known that solution of the stationary Dirac equation 
with finite basis sets leads to the 
emergence of the non-physical spurious states in the spectrum~\cite{bratzev1977,drake1981}. 
In our calculations, such states emerge and move upwards in the spectrum with 
the size of the basis set increase. We have checked, however, that the occurrence of such 
states does not affect the final results, since the transition probabilities are negligibly small for them. 

To solve the Eq.~\eqref{eq:tdde2d} we implement the time-dependent generalized pseudospectral 
(TDGPS) method~\cite{tong1997}, which was successfully employed in many previous
calculations~\cite{telnov2013,tumakov2017,telnov2018} for the Dirac equation.
The four-component function $\psi(r, \theta, t)$ is discretized on the two-dimensional 
GPS grid (see Ref.~\cite{telnov2013} for details) and propagated in time by 
the Crank-Nicolson (CN) propagation scheme~\cite{crank1947}:
\begin{equation}\label{eq:cn}
  \begin{split}
    \left[1 + \frac{i \Delta t}{2} \widetilde{H}\left(t + \frac{\Delta t}{2}\right) \right] \psi(r, \theta, t + \Delta t) &       \\
    = \left[1 - \frac{i \Delta t}{2} \widetilde{H}\left(t + \frac{\Delta t}{2}\right) \right] \psi(r, \theta, t),
  \end{split}
\end{equation}
with the initial condition set to be the ion ground state:
\begin{equation}\label{eq:ic}
\psi(r, \theta, t = 0) = \phi_{1s}(r, \theta).
\end{equation}
A set of the linear equations~\eqref{eq:cn} should be solved on each 
propagation step, which can be quite time-consuming , especially for extremely strong laser fields.  
However, the CN method allows the time step to be relatively large for obtaining 
converged results~\cite{telnov2018}. In addition, the matrix of the Hamiltonian $\widetilde{H}(t)$ is very sparse within 
the GPS discretization (given that almost always the size of the angular grid is much smaller that the radial one); in all our 
calculations the number of non-zeros in Hamiltonian matrix is around 3\%. These facts allow one to implement 
the scheme~\eqref{eq:cn} very efficiently using the existing libraries for the iterative solution of the linear equations 
with sparse matrices. In the present study we make use of the 
Intel\textsuperscript{\tiny\textregistered} MKL PARDISO~\cite{mkl} library.

\subsection{Direct PAD evaluation}
Having the full wavefunction at time $\tau > T$ after laser pulse is switched off, 
the momentum distribution of the photoelectrons can be obtained directly as
\begin{equation}\label{eq:directPAD}
\frac{d^2 P (\bm{k})}{dE d \Omega} = \frac{k}{c^2}E \sum_{\mu}|\braket{\Psi_{\bm{k} \, \mu}^{-}(\bm{r}) |
\Psi(\bm{r}, \tau)}|^2.
\end{equation}
Continuum wavefunction $\Psi_{\bm{k} \, \mu}^{-}(\bm{r})$ of the unperturbed Hamiltonian 
describes the electron state with the asymptotic momentum $\bm{k}$ 
(the corresponding energy is $E = c \sqrt{c^2 + k^2}$)
and the polarization $\mu$ in the far future. This function can be written as~\cite{rose1995,eichler2007, zaytsev2015}
\begin{eqnarray}
\Psi_{\bm{k} \, \mu}^{-}(\bm{r}) &=& \frac{1}{\sqrt{4 \pi}} \frac{c}{\sqrt{E k}} \sum_{\kappa m_j} C^{j \mu}_{l 0, 1/2 \mu} i^l
\nonumber\\
&\times& \sqrt{2 l + 1} e^{-i \delta_{\kappa}} D^j_{m_j \mu} (\bm{z} \rightarrow \bm{k}) \Psi_{E \kappa m_j} (\bm{r}).
\end{eqnarray}
Here $\kappa$ is the relativistic quantum number, $j = |\kappa| - 1/2$, $l = j + \frac{1}{2} {\rm sgn}(\kappa)$, $C^{j \mu}_{l 0, 1/2 \mu}$ is the Clebsch-Gordan coefficient, and 
$D^j_{m_j \mu} (\bm{z} \rightarrow \bm{k})$ is the Wigner matrix
rotating the $\bm{z}$ axis into the $\bm{k}$ direction~\cite{varshalovich1988}. 
The eigenfunctions of the Hamiltonian $H_0$ are normalized on the energy scale and 
take the form
\begin{equation}
\Psi_{E \kappa m_j} (\bm{r}) = \frac{1}{r} \begin{pmatrix} g_\kappa(r) \Omega_{\kappa m} (\bm{n}) \\ i f_\kappa(r) \Omega_{-\kappa m} (\bm{n}) \end{pmatrix}, 
\end{equation}
where $\Omega_{\kappa m} (\bm{n})$ is the spherical bispinor, and $\bm{n} = \bm{r} / r$. In the present work 
the radial functions $g_\kappa(r)$ and 
$f_\kappa(r)$ along with the scattering phase shifts $\delta_{\kappa}$ 
are obtained numerically using the modified \texttt{RADIAL} package~\cite{salvat1995}. The full continuum 
wavefunction is normalized as follows:
\begin{equation}
\left\langle \Psi_{\bm{k} \, \mu}^{-}(\bm{r}) | \Psi_{\bm{k'} \, \mu}^{-}(\bm{r}) \right \rangle = \delta(\bm{k} - \bm{k'}).
\end{equation}

\subsection{Relativistic Volkov functions within the dipole approximation}
In the original paper~\cite{volkov}, analytical solutions of the Dirac equation were 
introduced in case of the external field in the form of a plane wave. In the non-relativistic case Volkov 
functions within the dipole approximation can be built by propagation of the plane waves 
from $- \infty$ to the moment in time $\tau$ with the analytical propagator
\begin{equation}
U^{\rm V, \, NR}(\tau, - \infty) = \exp \left(-\frac{i}{2} \int_{- \infty}^{\tau} dt \left( \bm{k} + \bm{A}(t) \right)^2 \right);
\end{equation}
the analytical Volkov-type approximate 
solutions also can be built beyond the dipole approximation~\cite{boning2019}. However, the 
``dipole'' Volkov functions for the Dirac equations do not have closed-form expressions. It can be shown though, 
that the Volkov functions can be constructed with the numerical solution of the second-order 
differential equation for a scalar function~\cite{fradkin1991,gavrilov1996}. Here we will follow 
the procedure from Ref.~\cite{gavrilov1996}.

We start with the Dirac equation with the pure electric external field $\bm{F}(t) = - \frac{\partial}{\partial t} \bm{A}(t)$:
\begin{equation}
\left( i \beta \frac{\partial}{\partial t} - c \beta \boldsymbol{\alpha} \cdot \left( \bm{p} + \bm{A}(t) \right) - c^2 \right) \chi(\bm{r}, t) = 0.
\end{equation}
Substituting the solution $\chi$ in the form 
\begin{equation}
\chi(\bm{r}, t) = \left(i \beta \frac{\partial}{\partial t} - c \beta \boldsymbol{\alpha} \left( \bm{p} + \bm{A}(t) \right) - c^2 \right) \phi(\bm{r}, t),
\end{equation}
supposing $p_y = 0$ without loss of generality, 
and assuming the vector potential to be directed along the $z$ axis, we get:
\begin{eqnarray}\label{eq:diracSubst}
\frac{\partial^2 \phi(\bm{r}, t)}{\partial t^2} - i c F(t) \alpha_z 
\phi(\bm{r}, t) + c^2 p_x^2 \alpha_x^2 \phi(\bm{r}, t) \\ \nonumber
+ c^2 \left(p_z + A(t) \right)^2 \alpha_z^2 \phi(\bm{r}, t) -c^4 \phi(\bm{r}, 
t) = 0.
\end{eqnarray}
The solution of the Eq.~\eqref{eq:diracSubst} can be written in the form:
\begin{equation}\label{eq:diracFin}
\phi_{\bm{k} q s}(\bm{r}, t) = \phi_{\bm{k} s}(t) \exp \left(i \bm{k} \cdot \bm{r} \right) v_{s, \,q}, \; s = \pm 1, \, q = \pm 1, 
\end{equation}
where the constant orthonormal spinors $v_{s, \,q}$ ($v^{\dagger}_{s, \,q} v_{s, \,q'} = \delta_{q q'}$) satisfy the following equations:
\begin{equation}\label{eq:spinors}
\alpha_z v_{s, \, q} = s v_{s, \, q}; \; \beta \alpha_x v_{s, \, q} = i q v_{s, \, q}.
\end{equation}
From Eqs.~\eqref{eq:diracSubst}-\eqref{eq:spinors} one can obtain the following equation 
for the scalar function $\phi_{\bm{k} s}(t)$:
\begin{equation}\label{eq:scalarDirac}
\left( \frac{d^2}{dt^2} + c^2 (k_z + A(t))^2 + c^2 k_x^2 - i s c F(t) + c^4 
\right) \phi_{\bm{k} s}(t) = 0.
\end{equation}
The equation~\eqref{eq:scalarDirac} has two independent solutions, corresponding to the sign of the 
particle energy $\pm E = \pm c \sqrt{c^2 + k^2}$ (i.e. to the particle ($+$) and 
antiparticle ($-$))~\cite{fradkin1991} before the interaction with the external field is switched on. 
For the commonly used laser field parameters we can neglect the possible 
transitions between the negative and positive continua (i.e. pair creation in an electromagnetic field), and 
use only the solution corresponding to the positive continuum: 
\begin{equation}
i \frac{d}{dt} \phi_{\bm{k} s}(t) = E \phi_{\bm{k} s}(t), \, t \rightarrow - \infty.
\end{equation}
Also since the solutions 
with fixed $r$, $\bm{k}$ and different $s$ are linearly dependent~\cite{gavrilov1996}, only the independent 
functions $\phi_{\bm{k} 1}(t) \equiv \phi_{\bm{k}}(t)$ should be considered.

The ordinary differential equation~\eqref{eq:scalarDirac} can be efficiently solved numerically by the implicit Runge-Kutta method. The number of such 
equations, however, can be quite large depending on the desired momentum $\bm{k}$ resolution. Having the 
values $\phi_{\bm{k}}(t_j)$ with their derivatives, the set of 
the Volkov functions can be constructed as follows:
\begin{equation}\label{eq:volkov}
\chi^{\rm V}_{\bm{k} \, q}(t_j) = - \frac{1}{4 \pi^{3/2}} e^{i \bm{k} \cdot \bm{r}} \begin{pmatrix} 
-i q \alpha_{\bm{k} q} (t_j) \\ \alpha_{\bm{k} q} (t_j) \\ i q \beta_{\bm{k} q} (t_j) \\ 
\beta_{\bm{k} q} (t_j) \end{pmatrix},
\end{equation}
where 
\begin{equation}
\alpha_{\bm{k} q}(t) = i \frac{d}{dt}\phi_{\bm{k}}(t) + \left( c^2 -c (k_z + 
A(t)) - i q c k_x \right) \phi_{\bm{k}}(t),
\end{equation}
\begin{equation}
\beta_{\bm{k} q}(t) = i \frac{d}{dt}\phi_{\bm{k}}(t) + \left(-c^2 - c (k_z + 
A(t)) + i q c k_x \right) \phi_{\bm{k}}(t),
\end{equation}

Note that one should normalize the initial conditions for the Eq.~\eqref{eq:scalarDirac} to provide 
\begin{eqnarray}\label{eq:volkovNorm}
\frac{1}{2}[|\alpha_{\bm{k} q}(0)|^2 + |\beta_{\bm{k} q}(0)|^2] & = &  |\dot{\phi}_{\bm{k}}(0)|^2 + E^2 |\phi_{\bm{k}}(0)|^2 \\ \nonumber
&& + 2 c k_z {\rm Im}(\dot{\phi}_{\bm{k}}(0) \phi^*_{\bm{k}}(0)) = 1
\end{eqnarray}
in order to make the set of 
Volkov functions~\eqref{eq:volkov} orthonormal (for completeness the negative energy solutions should also be 
included in the set).

\subsection{Relativistic mask method}

Analogous to the non-relativistic mask method~\cite{chelkowski1998,tong2006,giovannini2012}, the whole coordinate space is divided into the inner (I) 
and outer (II) parts. Within the inner region, the wavefunction is propagated numerically with the method described 
in Sec.~\ref{subsec:tdde} with the full Hamiltonian. In the outer region, we neglect the interaction with the nucleus 
and make use of the Volkov states propagation in the momentum representation.

Let us consider the $j$'s step of the time propagation. At this step the wavefunction, which is a result of the propagation 
from the previous step in the inner region, should be divided in two parts by 
a smooth mask function $M(r)$ (``absorber''), which is equal to unity for $r < R$, and decreases to zero at some point $R_{\rm max}$:
\begin{equation}\label{eq:mask}
\Psi(\bm{r}, t_j) = \underbrace{M(r) \cdot \Psi(\bm{r}, t_j)}_{\Psi_{\rm I}} + \underbrace{(1 - M(r)) \cdot \Psi(\bm{r}, t_j)}_{\Psi_{\rm II}}.
\end{equation}
After that the numerical propagation according to the scheme~\eqref{eq:cn} continues only for the inner part $\Psi_{\rm I}$ 
step by step (with the procedure~\eqref{eq:mask} performed on each step) to some moment in 
time $\tau > T$. The value of $\tau$ should be large enough (typically, several optical cycles) 
so the absorber could ``capture'' the whole ionized wavepacket.
As a result, the normalization integral within the sphere of radius $R$
\begin{equation}
N(t) = \int_{r \leqslant R} d \bm{r} \left \langle \Psi(\bm{r}, t) | \Psi(\bm{r}, t) \right \rangle 
\end{equation}
is decreasing with time. Given that the value of $R$ is large enough, the quantity $P = 1 - N(\tau)$ after the end of 
the laser pulse $T$ can be interpreted as the ionization probability.

Note that the wavefunction splitting by the procedure~\eqref{eq:mask} does not allow the 
parts of the ionized wavepacket to move back from the outer to the inner region, so one should 
ensure that the value of $R$ at least exceeds the the electron excursion in the oscillating laser field. 
The final results should be checked in terms of convergence with respect to $R$. The value $R_{\rm max} - R$, i.e. the 
width of the absorber, can also influence the final results (see Ref.~\cite{giovannini2015} for the 
detailed discussion), so the convergence should be checked with respect to this quantity 
as well.

The absorbed outer parts ($\{\Psi_{\rm II}(t_j)\}$) of the wavefunction are used to 
calculate the following scalar products:
\begin{equation}\label{eq:cj}
C_j^{q}(\bm{k}) = \left\langle \chi^{\rm V}_{\bm{k} \, q}(t_j) | \Psi_{\rm II}(t_j) \right\rangle,
\end{equation}
where $\chi^{\rm V}_{\bm{k} \, q}(t_j)$ is the Volkov function~\eqref{eq:volkov}.

Finally, the differential ionization probability for the electrons
emitted with the momentum $\bm{k}$ into the unit energy and solid angle
intervals is evaluated as: 
\begin{equation}\label{eq:padmm}
\frac{d^2 P (\bm{k})}{dE d \Omega} = \frac{1}{c^2} E k \sum_{q} \left| \sum_j C_j^{q}(\bm{k}) \right|^2.
\end{equation}
The photoelectron energy spectrum can be obtained by integration of
PAD~\eqref{eq:padmm} over the angles:
\begin{equation}\label{eq:pes}
\frac{dP(E)}{dE} = \int \frac{dP(\bm{k})}{dE d \Omega} d \Omega.
\end{equation}
Then the additional integration of the spectrum~\eqref{eq:pes} over the emitted 
electron energy can be performed to obtain the ionization probability $P$:
\begin{equation}\label{eq:fullP}
P = \int_{0}^{\infty} \frac{dP(E)}{dE} dE.
\end{equation}
The comparison of the total ionization probability obtained with Eq.~\eqref{eq:fullP} 
with the same quantity evaluated as $1 - N(\tau)$ is used to control the accuracy of the results.

\section{Numerical examples and discussion}\label{sec:results}
To demonstrate the implementation of the relativistic mask 
method (RMM), we evaluate the PAD using two different ionization scenarios: 
multiphoton ionization of the Xe$^{53+}$ ion and over-the-barrier ionization 
of the hydrogen atom in superstrong laser fields. Since the chosen external 
fields in both cases possess high peak intensities and short wavelengths, 
we discuss the applicability of the dipole approximation first. As we mentioned in 
the Introduction, two conditions should be satisfied. The first one is 
smallness of the characteristic size of the target $r_{\rm ion}$ (measured as the 
mean distance of the electron from the nucleus in the ground state) with respect 
to the laser wavelength $\lambda$. The dipole approximation is expected to work
well if $\lambda/r_{\rm ion}\gg 1$. In our calculations (see the external field 
parameters below), $\lambda / r_{\rm ion} \sim$ 100 for the first example 
(Xe$^{53+}$), and $\lambda / r_{\rm ion} \sim$ 250 for the second one (H atom). 
The second condition assumes displacement of the classical 
electron due to 
the magnetic force in the direction perpendicular to that of the field 
polarization to be small compared to the electron wavepacket width at the 
moment of electron-target recollision. This condition is 
satisfied if $\Gamma_{\mathrm{R}} \ll 1$, where $\Gamma_{\mathrm{R}}$ is the 
Lorentz deflection parameter~\cite{palaniyappan2006}:
\begin{equation}
\Gamma_{\rm R} = 
\frac{\sqrt{I_{p}U_{p}^{3}}}{3\omega c^{2}}.
\label{eq:gamma}
\end{equation}
In Eq.~(\ref{eq:gamma}), $I_p$ is the ionization potential of the target and 
$U_{p}=F^{2}/(4\omega^{2})$ is the ponderomotive potential. In our 
calculations, 
$\Gamma_{\rm R} \sim 10^{-4}$ and $\Gamma_{\rm R} \sim 10^{-5}$ for the 
first and second case studies, respectively. At the end of this brief discussion, 
we may conclude that no significant difference between the present results and those 
obtained beyond the dipole approximation is expected for the parameters used in 
the calculations. Since the magnetic field effects scale as $v/c$ ($v$ is the 
characteristic electron velocity) while the other relativistic effects (like 
the mass correction) scale as $v^{2}/c^{2}$ at most, we do not expect large 
relativistic effects due to the external field (where $v\sim F/\omega$) either. 
However, the relativistic effects in the initial state of the heavy 
hydrogen-like ion (Xe$^{53+}$), such as a relativistic change of the ionization 
potential, are important and fully included in the solution of the Dirac 
equation.

\subsection{Multiphoton ionization of the H-like Xe ion}
As an example demonstrating applicability and 
performance of the RMM method, we evaluate the PAD for the H-like Xe ($Z$ = 54) 
ion after multiphoton above-threshold ionization (ATI) by a linearly-polarized 
laser pulse with the sin-squared envelope containing 20 optical cycles. The 
carrier wavelength is 0.1~nm (the corresponding photon energy is 455.63 a.u.) 
and the laser peak intensity is set to 8$\times$10$^{\textrm{23}}$ W/cm$^2$. The 
Keldysh parameter~\cite{keldysh1965} $\gamma$ is equal to 5.26, 
which corresponds to the multiphoton ionization regime.

\begin{figure}[t]
\includegraphics[width=1\linewidth]{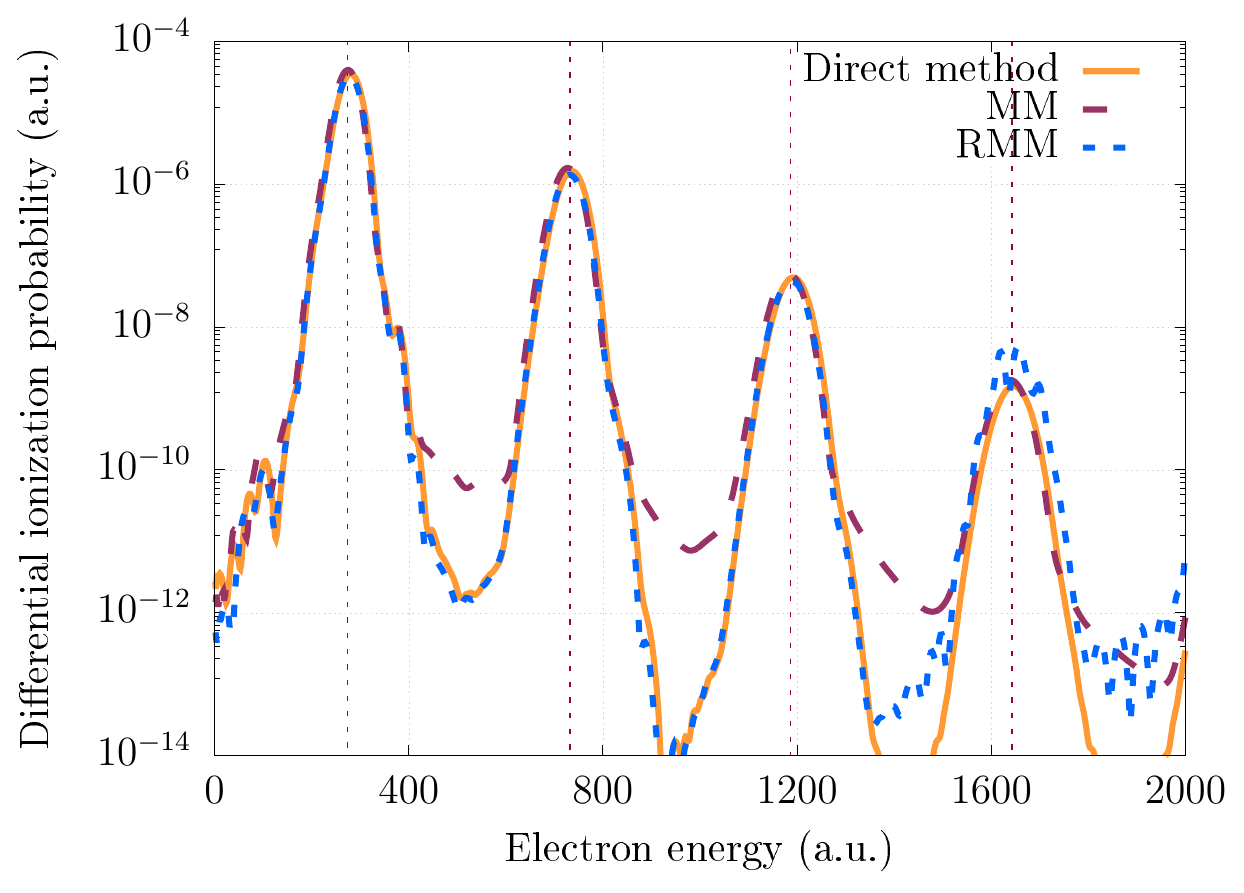}
\caption{Photoelectron spectra for the Xe$^{53+}$ ion exposed to a 
linearly-polarized laser pulse. Carrier wavelength is 0.1~nm, peak intensity is 
8$\times$10$^{\textrm{23}}$ W/cm$^2$. The pulse contains
20 optical cycles. Different curves correspond to the different way of 
the spectra evaluation.}
\label{fig:spectra_all}
\end{figure}

For comparison, we also show the results obtained with 
the usual non-relativistic mask method (MM) for the artificial ion with a value of $Z$ = 55.13, which 
gives the same ionization potential as the original ion and allows us to capture the 
major quantitative part of the relativistic effects in the considered process~\cite{vanne2012,ivanova2018}.

\begin{figure*}[t]
\subfigure{\label{fig:pad_direct}\includegraphics[height=4.1cm]{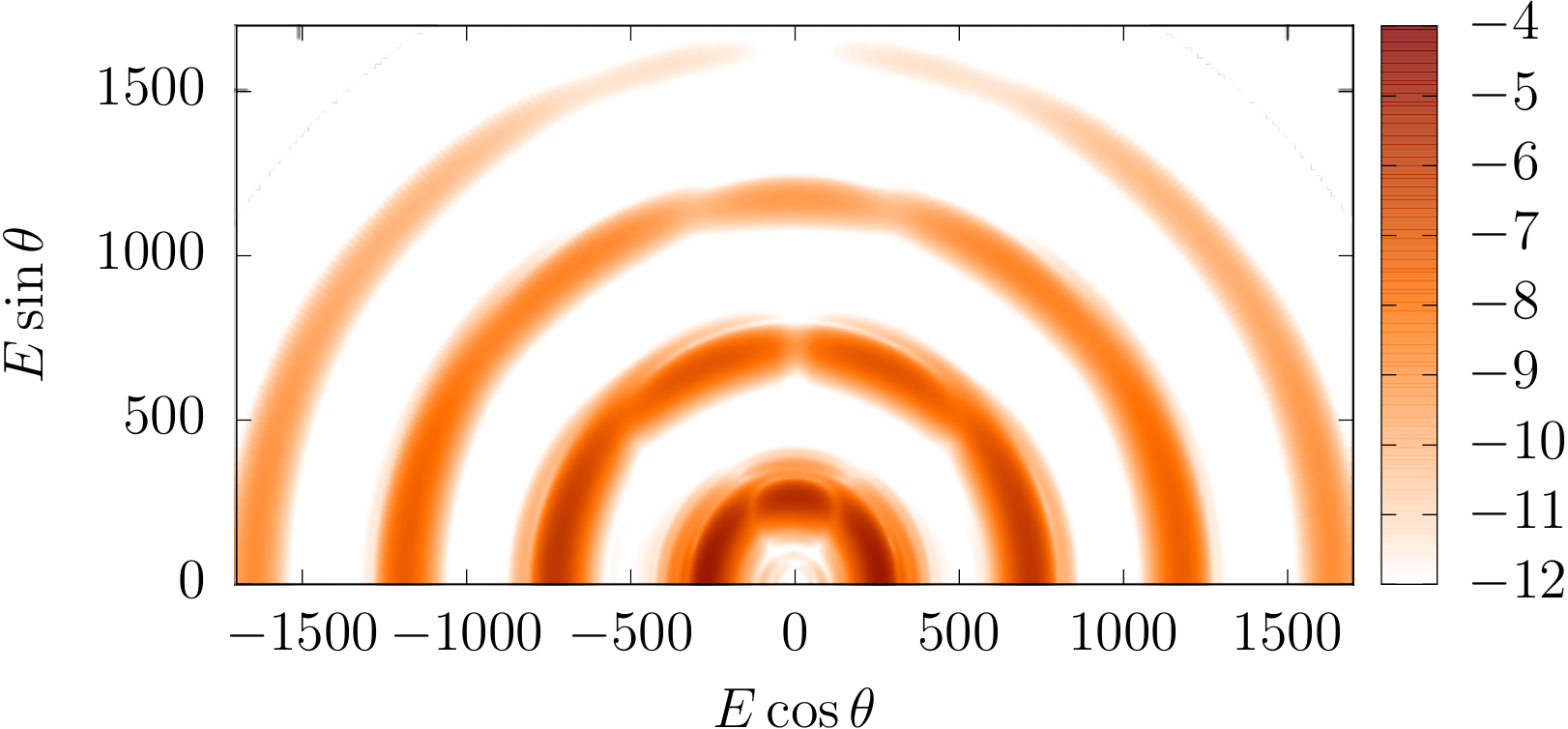}}
~~\subfigure{\label{fig:pad_mm}\includegraphics[height=4.1cm]{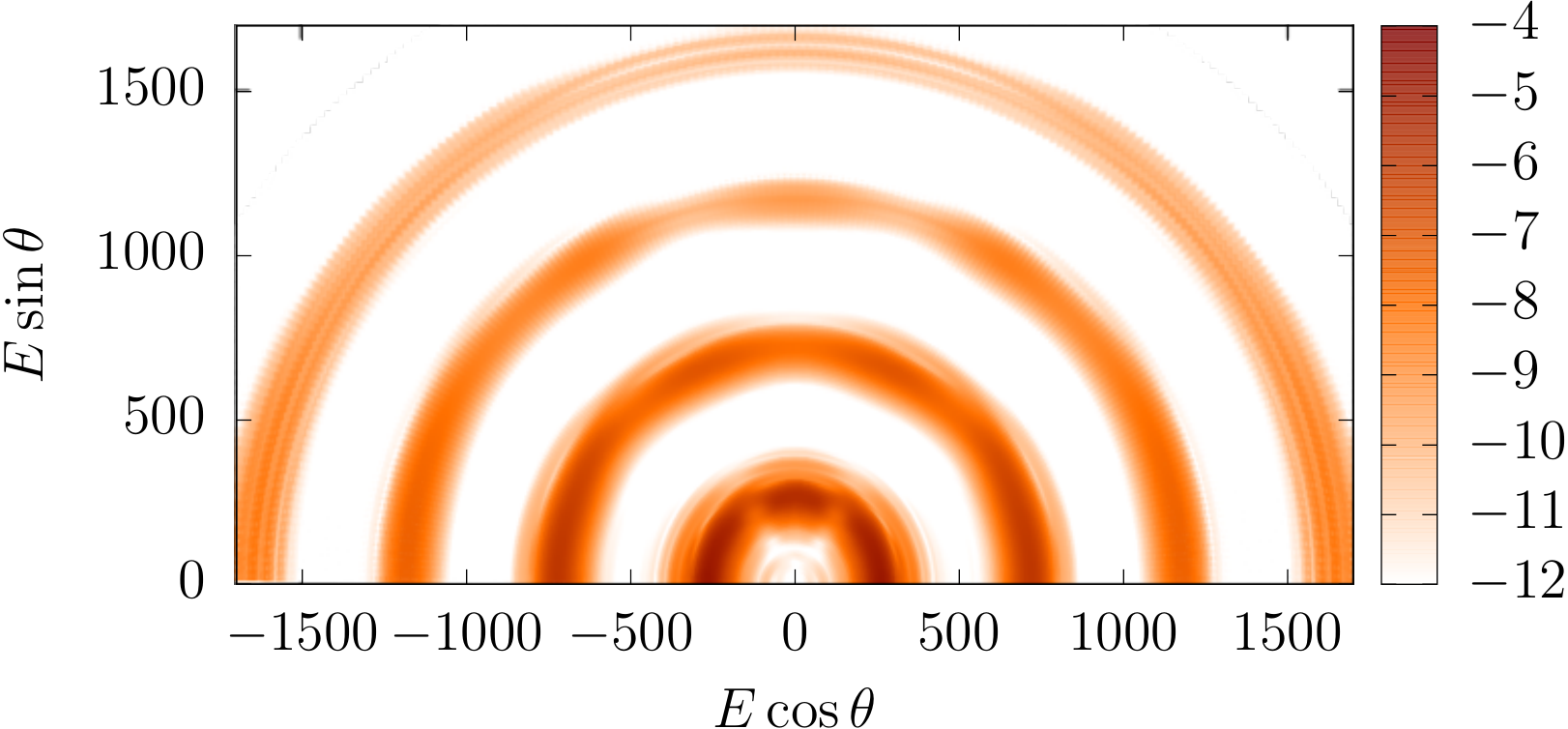}}
\caption{Two-dimensional energy-angular distributions for the Xe$^{53+}$ ion exposed 
to a linearly-polarized laser pulse calculated with the direct method (left) and 
with the relativistic mask method (right). Carrier wavelength is 0.1~nm, peak 
intensity is 8$\times$10$^{\textrm{23}}$ W/cm$^2$. The pulse contains
20 optical cycles. The PAD intensity scale is logarithmic and shown as
a color map.}
\label{fig:pad}
\end{figure*}
The photoelectron spectra calculated with different approaches are shown in 
Fig.~\ref{fig:spectra_all}. 
The vertical dashed lines represent the approximate positions of the peaks calculated 
as 
\begin{equation}\label{eq:peaks}
E_n = -I_p + n \omega - U_p,
\end{equation}
where the ionization potential $I_p$ is equal to 1519.47 a.u. 
for Xe$^{53+}$, and $n$ is the number of absorbed photons (for the 
process under consideration, $n \ge$ 4). One can see in 
Fig.~\ref{fig:spectra_all} that RMM successfully reproduces the spectra 
calculated with both the direct method and the non-relativistic treatment. 
However, the numerical simulation box with the linear dimension of 5 a.u. is 
sufficient for the RMM calculations while the direct method requires the box 
size twice as large.

Angular distributions calculated directly and with the RMM are depicted in 
Fig.~\ref{fig:pad}. The results are in 
good agreement here as well. Angular structure of the rings can be understood in terms of the dominant angular momentum 
of the photoelectrons~\cite{chen2006}, namely, the number 
of nodes is equal to the dominant value of the angular momentum in the continuum 
state. The first ring corresponds to the absorption of 4 photons, so it may 
contain contributions from the angular momenta 0, 2, and 4. The dominant
value of the angular momentum is equal to 2, as one can infer from the node 
structure of the first ring in Fig.~\ref{fig:pad}.

\subsection{Electron distributions after ionization of 
hydrogen atom in a superstrong laser field}
Here we study ionization of a hydrogen atom by a short 
linearly-polarized pulse of electromagnetic radiation with the carrier 
wavelength 13 nm (photon energy 3.5 a.u.). The pulse envelope has a 
$\sin^{2}$ shape with the peak intensity 3.5$\times 10^{18}$ W/cm$^2$ and 
contains 5 optical cycles. Since the ionization potential of the hydrogen atom 
is equal to 0.5 a.u., we have an ionization regime where both the photon 
energy essentially (about 7 times) exceeds the ionization potential and the peak 
value of the external field exceeds by far (about 19 times) the Coulomb force 
from the nucleus on the first Bohr orbit.

\begin{figure}[t]
\includegraphics[width=1\linewidth]{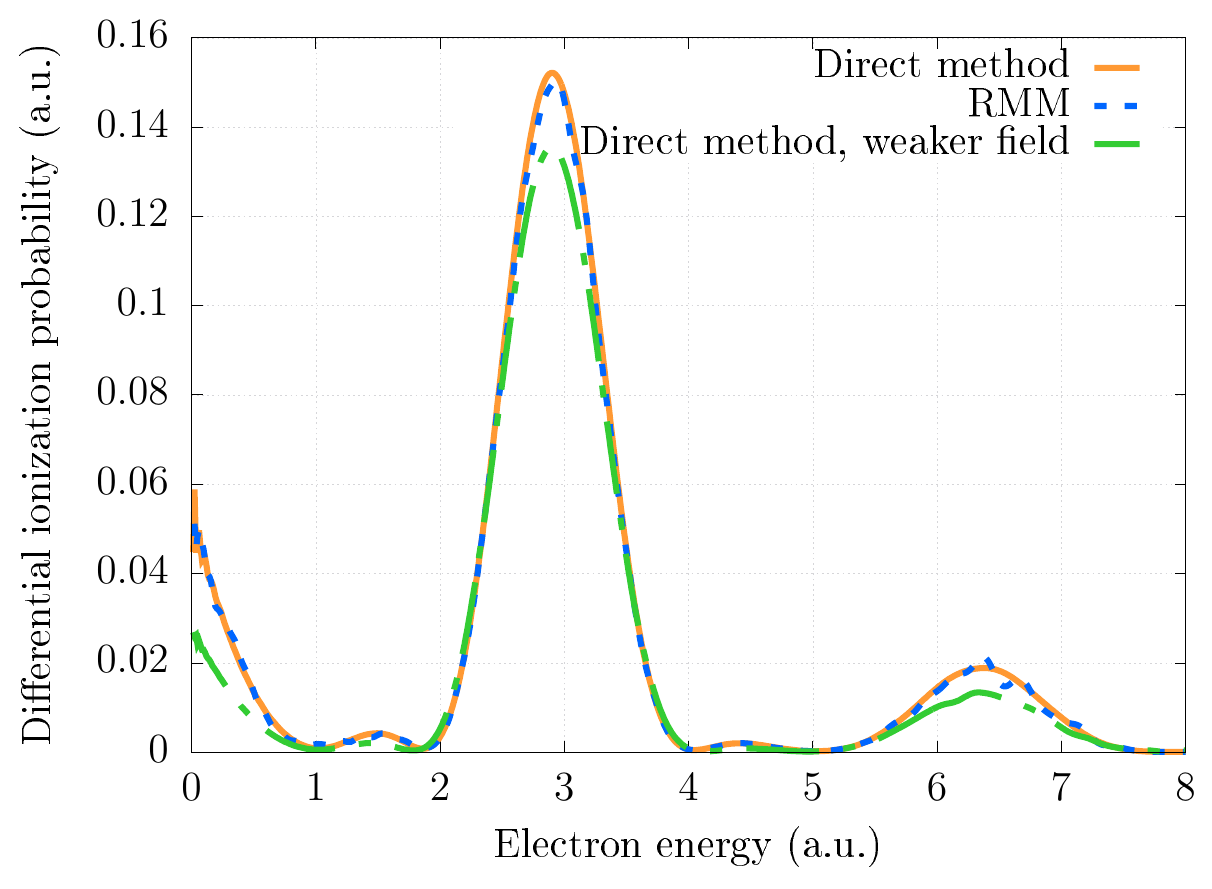}
\caption{Photoelectron spectra for the hydrogen atom  
exposed to a linearly-polarized laser pulse with the duration of 5 optical 
cycles. The carrier wavelength is 13~nm and the peak intensity is 3.5$\times 
10^{18}$ W/cm$^2$ (except for ``weaker field'' curve, which corresponds to the 
peak intensity of 2.3$\times 10^{18}$ W/cm$^2$). The methods used to evaluate 
the spectra are as labeled.}
\label{fig:spectra_all_h}
\end{figure}

The results of our calculations are presented in Fig.~\ref{fig:spectra_all_h}.  
One can clearly see two ATI peaks, the first one centered 
around 2.8~a.u., and the second one around 6.3~a.u. The RMM and direct method 
results agree very well. To capture the ``slow'' electrons with the mask 
method, one should propagate the wavefunction for a long time after the 
interaction with the laser. We avoid this by calculating the low-energy 
structure with the direct method, thus a hybrid method is 
eventually used with the two different techniques applied on different energy 
ranges.

Note that the ponderomotive potential is very large for this superstrong field: 
$U_{p} =$2~a.u. According to Eq.~\eqref{eq:peaks}, one should observe a 
large shift (comparable with the peak spacing) in the 
positions of the ATI peaks. Clearly this is not the case, as one can see in 
Fig.~\ref{fig:spectra_all_h}. Moreover, the ATI peak positions have only very 
weak dependence on the laser field intensity (the ``weaker field'' curve in 
Fig.~\ref{fig:spectra_all_h} corresponding to the peak intensity of 
2.3$\times$10$^{\textrm{18}}$ W/cm$^2$ is  slightly shifted \textit{to the 
left}), again in contradiction with 
Eq.~\eqref{eq:peaks}. A breakdown of Eq.~\eqref{eq:peaks} in superstrong laser 
fields, however, can be easily understood. The formula  \eqref{eq:peaks} for the 
positions of the ATI peaks is based on a simple and intuitive picture: while the 
continuum and weakly bound states are strongly perturbed by the external 
oscillating field, and their energies are shifted by the mean oscillation energy 
$U_{p}$, the tightly bound ground state is perturbed only weakly, and its energy 
remains unchanged (if one neglects the relatively small ac Stark shift). 
Therefore the ionization potential is effectively increased by $U_{p}$, and one 
can observe the ATI peaks moving towards lower energies and even change in the 
minimum number of photons required for ionization (peak switching  
\cite{kruit1983}), as the intensity of the laser field becomes higher. When the 
field becomes superstrong, however, this picture changes dramatically. In a
superstrong external field, not only the continuum and weakly bound states but 
also tightly bound states, including the ground sate, are strongly perturbed by 
the field. Now all the states, bound and continuum, are shifted by the same 
amount of energy, $U_{p}$. Hence the ionization potential does not change 
compared to the case of the weak external field where both $U_{p}$ and ac Stark 
shift are small. Therefore in a superstrong field one can see the ATI 
peaks approximately at the same positions as in a relatively weak laser field, 
that is, not shifted by the ponderomotive potential.

\begin{figure}[t]
\includegraphics[width=1\linewidth]{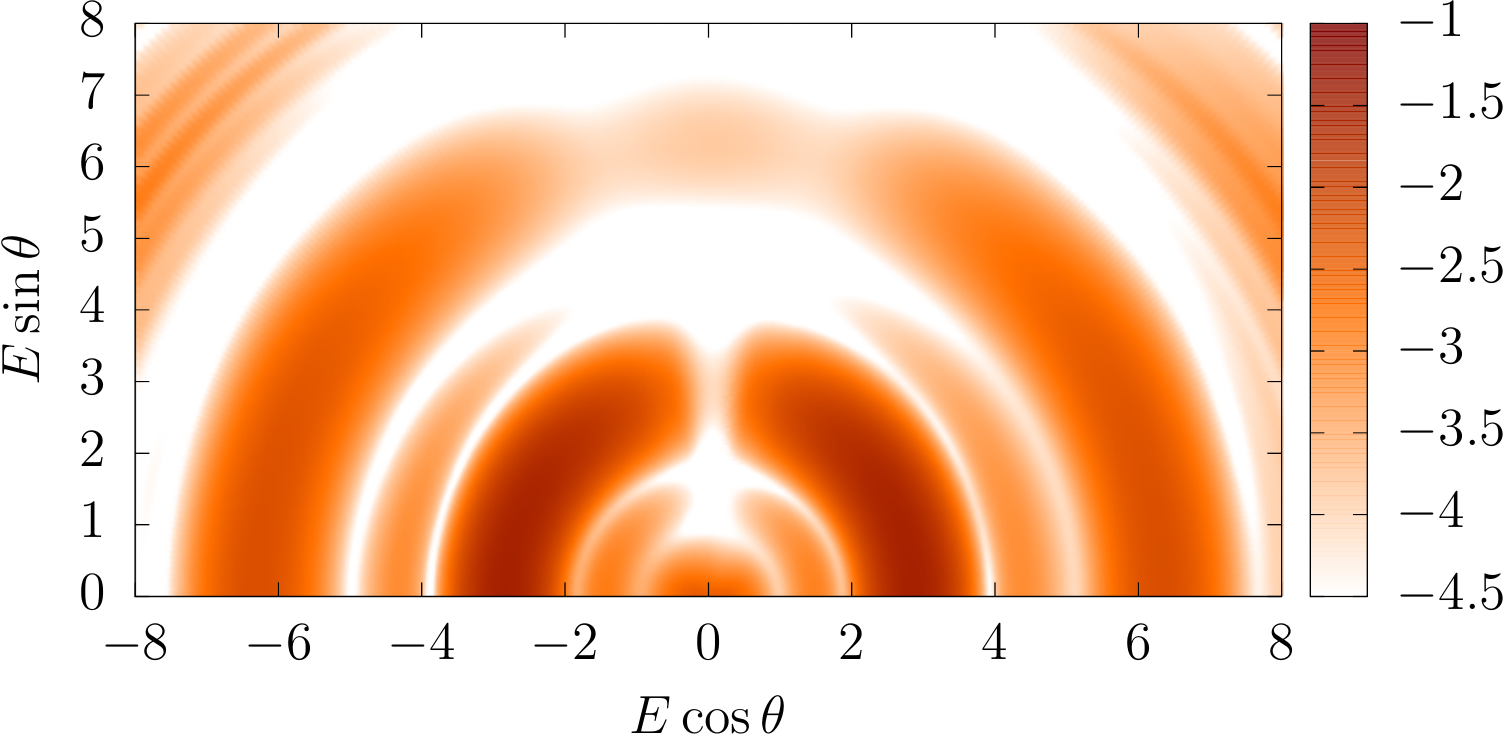}
\caption{Two-dimensional energy-angle distribution for the hydrogen atom exposed to a 
linearly-polarized laser pulse. Carrier wavelength is 13~nm, peak intensity is 
3.5$\times$10$^{\textrm{18}}$ W/cm$^2$. The pulse contains
5 optical cycles. The PAD intensity scale is logarithmic and shown as
a color map.}
\label{fig:pad_direct_h}
\end{figure}

The two-dimensional energy--angle spectra calculated with 
the direct method (and reproduced with the mask method)
are presented in Fig.~\ref{fig:pad_direct_h}. One can see slight 
forward-backward asymmetry, what is usually the case for 
the short pulses. The main one- and two-photon absorption 
rings are accompanied by weaker satellites. This multiring oscillatory 
structure within the ATI peak is due to interference of the electronic signal 
coming from the leading and trailing edges of the laser 
pulse \cite{telnov1995}.

\section{Conclusion}\label{sec:summary}
In this contribution, we have extended the mask (or 
wavefunction-splitting) method for evaluation of the photoelectron energy 
spectra and angular distributions to the relativistic domain within the dipole 
approximation. Despite the dipole approximation lacks analytical closed-form 
expressions for the relativistic Volkov states, a set of uncoupled second-order 
ordinary differential equations for a scalar function can be efficiently solved 
to construct such states numerically. Quantum electrodynamics 
effects are neglected in the present 
study because the electromagnetic fields used in the calculations are not 
strong enough to make these effects significant for the processes under consideration.

To demonstrate the implementation of the method, we have 
performed two case studies, both within the dipole approximation: multiphoton 
above-threshold ionization of the hydrogen-like Xe$^{53+}$ ion and ionization of 
the hydrogen atom in a superstrong electromagnetic field. The 
ionization regimes in these two cases are considerably different. Nonetheless, 
in both cases the relativistic mask method is able to reproduce the electron 
energy and angular distributions calculated with the direct approach, that is by 
projecting the final wavefunction onto the continuum states of the target ion. 
Although the direct method is straightforward to implement, it has evident 
limitations. This method may have serious difficulties in evaluation of the 
spectra of fast photoelectrons, which may leave the simulation box before the 
laser field is switched off. For the same reason, longer pulses are not 
well-suited for the direct method either. Certainly, these difficulties can be avoided 
by expansion of the simulation box, but this will result in more demanding and 
heavy computations. In this respect, the relativistic mask method is more robust 
and efficient since it can catch the fast electrons ``on the fly'' before they 
leave the box, so a huge simulation box is not required. 
Although in the present formulation the relativistic mask 
method is restricted to the dipole approximation, we believe it can find its 
applications, for example, in comparison with the full nondipole calculations 
for estimation of the nondipole effects.

%
%
\section{Authors contributions}
D.~A.~Tumakov carried out the numerical calculations. All the authors were 
involved in discussions and preparation of the manuscript. All the authors have 
read and approved the final manuscript.

\acknowledgments
This work was supported by Russian Foundation for Basic 
Research (Grant No. 20-02-00199). D.~A.~Tumakov acknowledges the support from TU 
Dresden (DAAD-Programm Ostpartnerschaften). The calculations were performed at 
the Computing Center of SPbSU Research Park.

\bibliography{rmm}

\end{document}